\documentclass[12pt]{scrartcl}

\usepackage[centertags]{amsmath}
\usepackage{amssymb}
\usepackage{hyperref}
\hypersetup{
    bookmarks=false,      
    pdfstartview={FitH},  
    colorlinks=true,      
    linkcolor=black,      
    citecolor=blue,       
    filecolor=blue,       
    urlcolor=blue         
}

\usepackage{graphicx}
\usepackage{varioref}
\usepackage{paralist}
\usepackage{subfigure}

\labelformat{section}{Section #1} 
\labelformat{subsection}{Section #1} 
\labelformat{equation}{Eq.~(#1)} 
\labelformat{figure}{Fig.~#1} 
\labelformat{subfigure}{Fig.~\thefigure#1} 

\setkomafont{subsection}{\normalcolor\bfseries}
\addtokomafont{caption}{\footnotesize}

\newcommand{\vev}{{\small{\textit{vev}}}}

\newcommand{\MS}{\ensuremath{\overline{\mathrm{MS}}}}
\newcommand{\FP}{\ensuremath{\lambda_{\mathrm{FP}}}}
\newcommand{\SU}[1]{\ensuremath{\mathrm{SU}(#1)}}

\begin{document}

\pagenumbering{roman}

\titlehead{\hfill LPSC-11192}
\title{\Large Implications of the Stability and Triviality Bounds on the Standard Model with Three and Four Chiral Generations}
\author{Ak\i{}n Wingerter\footnote{akin@lpsc.in2p3.fr}\\
{\normalsize \it Laboratoire de Physique Subatomique et de Cosmologie}\\[-1ex]
{\normalsize \it UJF Grenoble 1, CNRS/IN2P3, INPG}\\[-1ex]
{\normalsize \it 53 Avenue des Martyrs, F-38026 Grenoble, France}\\
}
\date{}
\dedication{{\bfseries Abstract}\\[2ex]
\begin{minipage}[b]{0.9\linewidth} 
\small  We revisit the stability and triviality bounds on the Higgs boson mass in the context of the Standard Model with three and four generations (SM3 and SM4, respectively). In light of the recent results from LHC the triviality bound in the SM3 has now become obsolete, and the stability bound implies for a Higgs mass of e.g.~$m_H=115$ GeV the onset of new physics before $\Lambda=650$ TeV, whereas there are no limits for $m_H\geq133$ GeV. For the SM4, the stability and triviality curves intersect and bound a finite region. As a consequence, the fourth generation fermions place stringent theoretical limits on the Higgs mass, and there is a maximal scale beyond which the theory cannot be perturbatively valid. We find that the Higgs mass cannot exceed 700 GeV for any values of the fourth generation fermion masses. Turning the argument around, the absence of a Higgs signal for $m_H\leq600$ GeV excludes a fourth generation with quark masses below 300 GeV and lepton masses below 350 GeV. In particular, the quark bounds also hold for the small mixing scenarios for which the direct limits from Tevatron and LHC are not applicable, and the lepton bounds we obtain are stronger than the collider limits. If a Higgs boson lighter than 700 GeV is not observed, a fourth generation of chiral fermions with perturbative Yukawa couplings will be conclusively excluded for the full range of parameters.
\end{minipage} }

\maketitle[0]
\thispagestyle{empty}
\newpage

\pagenumbering{arabic}

\clearpage
\newpage
\section{Introduction}

Recent years have seen a renaissance of the idea of a fourth generation of chiral fermions. One reason is the realization that the constraints are by far not as stringent as they were thought to be \cite{Alwall:2006bx,Holdom:2009rf}. Another is that an extra generation of fermions would allow for higher Higgs masses and thereby ease the tension between the lower limit set by LEP \cite{Barate:2003sz} and the value favored by electroweak precision measurements \cite{aleph:2010vi}.

\medskip

In the Standard Model (SM), the stability of the vacuum gives a lower bound on the Higgs mass \cite{Cabibbo:1979ay}. Similarly, requiring that the Higgs self-coupling not blow up at high scales limits $m_H$ from above \cite{Lindner:1985uk}. The consequences of the stability and triviality bounds (as the latter one is called) in the SM have been analyzed in great detail (see e.g.~Ref.~\cite{Djouadi:2005gi} and references therein). Here we extend this analysis to the case of the Standard Model with four chiral generations (SM4).
 
\medskip

The stability and triviality bounds depend sensitively on the fourth generation quark masses $m_{t'}$ and $m_{b'}$, and to a lesser extent on $m_{\tau'}$ and $m_{\nu'}$. We find that for fourth generation fermion masses consistent with experiment, the allowed range for the Higgs mass is roughly from 400 to 700 GeV, but grows smaller very fast with increasing scale up to which we assume the SM4 to be valid. The most striking difference as compared to the Standard Model with three chiral generations (SM3) is that there is no ``corridor'' of allowed Higgs masses where the theory is valid up to some grand unification scale; rather, the stability and triviality curves intersect and determine the scale where new physics must
necessarily take over.

\medskip

The ATLAS and CMS experiments with an integrated luminosity of more than 1 fb${}^{-1}$ at $\sqrt{s}=7$ TeV have already ruled out the Higgs boson in the mass range of 120-600 GeV in the context of the SM4 \cite{ATLAS-CONF-2011-135,CMS-HIG-11-011}. The enhanced sensitivity of the LHC experiments to a SM4 Higgs boson is mainly due to the extra fermions contributing to Higgs production by gluon fusion and leads to a ``supervisible'' Higgs in contrast to most beyond the Standard Model scenarios. If a Higgs boson lighter than roughly 700 GeV is not observed, this will be at odds with the existence of a fourth chiral generation of fermions, since the triviality and stability bounds do not allow for a higher Higgs mass for any choice of the fourth generation fermion masses, if the Yukawa couplings are to remain perturbative. The current Higgs exclusion limit of 120-600 GeV can be used to set limits on the fourth generation quark and lepton masses that are competitive with and in some cases better than the corresponding limits from collider experiments.

\medskip

The outline of the paper is as follows. In \ref{sec:expconstraints} we critically discuss the experimental limits on the fourth generation fermion masses and the SM4 Higgs mass. As we will see, for certain values of the mixing angles and masses, the collider limits can be significantly weakened. In \ref{sec:stability-and-triviality} we explain in detail the steps we followed to obtain the stability and triviality bounds. In \ref{sec:results} we present our main results. In \ref{sec:results_SM3} we update the stability and triviality bounds for the Standard Model with three generations using the most recent experimental values of the strong coupling constant $\alpha_s$ and the top quark $m_t$, and we take care of several typos and misprints in the 2-loop renormalization group equations (RGEs) that have entered some of the previous calculations. We also comment on some small numerical disagreement we have with Ref.~\cite{Hambye:1996wb}. In \ref{sec:results_SM4} we generalize the stability and triviality bounds to the case of the Standard Model with four chiral generations. Our work improves upon previous results in the literature in that we use the full 2-loop RGEs with generalized 1-loop matching corrections for the top quark and the Higgs boson, and furthermore include neutrino masses and quark mixing which cannot be neglected in the presence of a fourth generation. In \ref{sec:theolimits} we discuss our theoretical constraints on the fourth generation fermion masses. Finally, in \ref{sec:conclusions} we present our conclusions. Some lengthy, but important details of this work have been relegated to the appendices.

\section{Experimental Constraints}
\label{sec:expconstraints}

\subsection{Direct Limits on Fourth Generation Fermion Masses}
\label{sec:directlimits}

The most stringent bounds on the fourth generation fermion masses come from the Tevatron \cite{Aaltonen:2011tq,Aaltonen:2011vr} and LEP \cite{Nakamura:2010zzi} experiments:
\begin{equation}
m_{t'}>358 \text{ GeV}, \enspace m_{b'}>372 \text{ GeV}, \enspace m_{\tau'}>100.8 \text{ GeV}, \enspace m_{\nu'}>80.5 \text{ GeV} \enspace \text{at }95\% \text{ C.L.}
\label{eq:explimitsfermions}
\end{equation}  
Preliminary results from the LHC experiments indicate that $m_{t'}>450$ GeV \cite{CMS-PAS-EXO-11-051} and $m_{b'}>495$ GeV \cite{CMS-PAS-EXO-11-036}.

\medskip

Ref.~\cite{Hung:2007ak} points out that the limits obtained by Tevatron can be considerably weakened:
\begin{inparaenum}[(i)]
\item If $t'\rightarrow W\,b'$ is kinematically allowed, it may dominate over the decay mode $t'\rightarrow W\,q$ with $q=d,s,b$ that was used in the search for $t'$.
\item $t'$ may be long-lived and decay outside the silicon vertex detector in a region where the searches for stable quarks do not yet apply.
\end{inparaenum}
In the latter case, there may be no limits on $m_{t'}$ at all. Both scenarios sensitively depend on $m_{b'}$ and the mixing angle between the third and fourth generations. The comments of Ref.~\cite{Hung:2007ak} concern earlier Tevatron results, but the more recent analysis on $m_{t'}$ \cite{Aaltonen:2011tq} makes the same set of assumptions. In particular, Ref.~\cite{Aaltonen:2011tq} states that electroweak precision measurements disfavor $m_{t'}-m_{b'} < m_W$ which would forbid $t'\rightarrow W\,b'$. However, Fig.~(13) of Ref.~\cite{Baak:2011ze} shows that electroweak precision data can easily accommodate a mass difference $m_{t'}-m_{b'} \sim m_W$ at 95\% C.L.~as long as $m_H\gtrsim350$ GeV. Regarding the second scenario, one may rightly deem it very improbable that the mixing angle is so tiny but yet non-vanishing to allow $t'$ to decay outside the silicon vertex detector. Contrary to the case of neutrino mixing, however, in the quark sector there is a clear tendency of the heavier families to mix less and less with the lighter ones. As a consequence, for a comprehensive analysis the small mixing angle scenario should be taken into consideration. 

\medskip

The analysis on $m_{b'}$ \cite{Aaltonen:2011vr} assumes that the branching ratio for $b'\rightarrow W\,t$ is 100\%. If one allows for non-zero mixing between the fourth and the first three generations and assumes that $m_{t'}>m_{b'}$, the lower limit on $m_{b'}$ decreases by as much as 10-20\% \cite{Flacco:2010rg}. As before, for a small region in parameter space, the $b'$ may decay outside the silicon vertex detector detector and thus escape detection \cite{Hung:2007ak}.

\medskip

The bound on $m_{\tau'}$ assumes that $\tau'$ can decay to at least one of the light neutrinos of the first three generations. Similarly, the bound on $m_{\nu'}$ assumes that the neutrino has Majorana couplings to at least one of the charged leptons of the first three generations; in case of a Dirac neutrino, the lower bounds are slightly higher. In light of the neutrino mixing pattern that has emerged over the past 10 years, we consider both assumptions to be safe, since we are considering a \textit{sequential} fourth generation. In the unfavorable case that all mixing angles between the first three generations and the fourth one vanish, we can still rely on the lower bound $m_{\nu'}>45$ GeV as long as $\nu'$ couples to the $Z$ boson. If this should not be true, one needs to reconsider whether $\nu'$ can rightly be called the ``neutrino'' in a sequential fourth generation.

\subsection{Indirect Limits on Fourth Generation Fermion Masses}

The electroweak precision measurements can accommodate a wide range of fourth generation fermion masses and are more sensitive to the mass splittings $\left|m_{t'}-m_{b'}\right|$ and $\left|m_{\tau'}-m_{\nu'}\right|$ than to the individual masses $m_{t'}$, $m_{b'}$, $m_{\tau'}$, $m_{\nu'}$ \cite{Baak:2011ze}. Depending on the Higgs mass $m_H$ and the mass difference between the leptons, the splitting between the quark masses can be as large as 90 GeV (see Fig.~(13) of Ref.~\cite{Baak:2011ze}). In the lepton sector, $m_{\tau'}>m_{\nu'}$ is preferred, and this effect becomes more pronounced for larger Higgs masses. For the quarks, there is no preference in the hierarchy. The analysis assumes that there is no mixing between the first three generations and the fourth one.

\medskip

An upper limit on the fourth generation fermion masses (the so-called unitarity bound) can be inferred by considering the partial wave expansion of their scattering amplitudes: $m_{t'}$, $m_{b'}$ $\lesssim500$ GeV and $m_{\tau'}$, $m_{\nu'}$ $\lesssim1000$ GeV \cite{Chanowitz:1978mv}. Note, though, that these bounds do not constitute strict limits, but rather indicate when strong dynamics should take over, if the bounds were violated. More stringent limits can be derived by considering the RG running of the fourth generation fermion Yukawa couplings and demanding that they remain perturbative up to a certain scale. We will indicate these ``perturbativity limits'' in the plots that we will present in \ref{sec:results}.

\medskip

The mass bounds on the fourth generation fermions depend to some extent on the mixing between the fourth and the first three generations (see remarks in \ref{sec:directlimits}). The measured values of the 3-by-3 CKM matrix and flavor changing neutral currents (FCNCs) constrain the mixing parameters of the full 4-by-4 matrix, but the bounds are comparatively weak: $\theta^{\textrm{CKM}}_{14}<3^\circ$, $\theta^{\textrm{CKM}}_{24}<8^\circ$, $\theta^{\textrm{CKM}}_{34}<42^\circ$ \cite{Bobrowski:2009ng}. Constraints from electroweak precision observables give $\theta^{\textrm{CKM}}_{34}\lesssim13^\circ$, i.e.~mixing between the third and fourth generations is at most of order the Cabibbo angle \cite{Chanowitz:2009mz}.

\subsection{Direct and Indirect Limits on the Higgs Mass}
\label{sec:higgslimits}

For the Standard Model, the LEP experiments have set a lower limit of $m_H>114.4$ GeV \cite{Barate:2003sz}. The ATLAS and CMS experiments exclude at 95\% C.L.~a Higgs lying in the ranges 146-232 GeV, 256-282 GeV, 296-466 GeV \cite{ATLAS-CONF-2011-135}, and 145-216 GeV, 226-288 GeV, 310-400 GeV \cite{CMS-TALK-LP2011}, respectively. The Tevatron gives no further constraints.

\medskip

In the presence of a fourth generation, the Higgs production cross section at the LHC increases by a factor of 4 to 9 \cite{tanaka:fourthgenerationmodel}. More relevant for the observation of the Higgs boson is the product of its production cross section with the modified \cite{tanaka:fourthgenerationmodel} branching ratios of its decay modes. Fig.~(5) of Ref.~\cite{Ruan:2011qg} shows that this product is 4-8 times larger than its corresponding value in the SM3. As a consequence, the LHC is much more sensitive to searches for a SM4 Higgs boson and can exclude it at 95\% C.L.~in the mass range from 120-600 GeV \cite{ATLAS-CONF-2011-135,CMS-HIG-11-011}. Generally, the analyses assume that the extra fermions are ultra-heavy in order to place the most conservative limits (see again Fig.~(5) of Ref.~\cite{Ruan:2011qg}). Also, note that 600 GeV is the maximum Higgs mass covered by the current analyses. In our context this is important, since depending on the masses of the fourth generation fermions, the Higgs boson in the SM4 can be as heavy as 700 GeV (see \ref{sec:results}). 

\medskip

In the case of the SM4, the fourth generation neutrino is necessarily heavy (see \vref{eq:explimitsfermions}) and the invisible decay mode $H\rightarrow\nu'\nu'$ has to be taken into consideration. However, Fig.~1 of Ref.~\cite{Rozanov:2010xi} shows that for $m_{\nu'}\simeq80$ GeV the dominant branching ratio $H\rightarrow WW$ is an order of magnitude larger and is not affected by the presence of the heavy neutrino. In the unlikely case that $m_{\nu'}\ll80.5$ GeV and $m_{\nu'}>45$ GeV (see our remarks at the end of \ref{sec:directlimits}) the decay $H\rightarrow\nu'\nu'$ cannot be neglected in the regime of small Higgs masses $m_H\lesssim200$ and the search strategy needs to be modified \cite{Rozanov:2010xi}. For $H\rightarrow\tau'\tau'$ the same arguments apply as for the neutrinos. The case of the quarks, however, is more problematic. For large Higgs masses $m_H\gtrsim400$ GeV and not too heavy quarks $m_{t'},m_{b'}\lesssim300$ GeV, the branching ratio of $H\rightarrow t't'$ can be competitive with $H\rightarrow Z Z$ from which the Higgs searches in that mass region derive their main sensitivity. Whether this can have an impact on the experimental analyses and as a consequence on the SM4 Higgs exclusion limits is a question that may need some consideration. Most analyses assume that the quark masses are of order $600$ GeV or $\infty$ \cite{ATLAS-CONF-2011-135,CMS-HIG-11-011,tanaka:fourthgenerationmodel}.

\medskip

The requirement of perturbative unitarity of gauge boson scattering gives the \textit{tree-level} unitarity bound $m_H\lesssim1$ TeV \cite{Lee:1977yc} that is valid both for the SM3 and the SM4. In the case of the SM3, precision electroweak data gives $m_H<185$ GeV at 95\% C.L.~\cite{aleph:2010vi}. For the SM4, Higgs masses as large as 900 GeV can be compatible with all electroweak precision measurements as long as the fourth generation fermion masses are adjusted accordingly \cite{Baak:2011ze}.

\section{The Stability and Triviality Bounds}
\label{sec:stability-and-triviality}

For deriving the stability and triviality bounds for the SM4, we mainly follow the discussion in Ref.~\cite{Hambye:1996wb}. We use the full 2-loop renormalization group equations (RGEs) \cite{Machacek:1983tz,Machacek:1983fi,Machacek:1984zw,Arason:1991ic,Ford:1992mv,Luo:2002ey} including neutrinos \cite{Pirogov:1998tj} and use $\overline{m}_b=4.19$ GeV (\MS{} mass) \cite{Nakamura:2010zzi} and $m_t=173.3$ GeV (pole mass) \cite{:1900yx}. The electroweak gauge coupling constants $g$, $g'$ are calculated from $\alpha^{-1}_\mathrm{em}=127.90$ and $\sin^2\theta_w=0.2315$, and we take $\alpha_s=0.1184$ \cite{Nakamura:2010zzi} (all in $\overline{\text{MS}}$ scheme at $M_Z=91.1876$ GeV), and finally $v=246.22$ GeV for the vacuum expectation value (\vev{}) of the Higgs. Some issues regarding the RGEs are discussed in \ref{sec:RGESM4}.

\medskip

Beyond tree level, the relation between the $\overline{\text{MS}}$ quartic Higgs coupling $\lambda$ and the physical Higgs mass $m_H$ will receive radiative corrections, and the same is true for the relation between the $\overline{\text{MS}}$ top Yukawa coupling $y_t$ and the pole mass $m_t$:
\begin{equation}
\lambda(\mu) = \left.\frac{m_H^2}{v^2}\left(1+\delta_H(\mu)\right)\right|_{\mu=m_H}, \quad \left.y_t(\mu) = \frac{\sqrt{2}m_t}{v}\left(1+\delta_t(\mu)\right)\right|_{\mu=m_t}
\label{eq:higgstopmatching}
\end{equation}
We have generalized the so-called Higgs and top matching corrections \ref{eq:higgstopmatching} to include corrections from the fourth generation fermions. The details can be found in \ref{sec:higgsmatchingconditions} and \ref{sec:topmatchingconditions}. For a discussion of the appropriate choice for the matching scales ($\mu=m_H$ and $\mu=m_t$, respectively), see Ref.~\cite{Hambye:1996wb}.

\medskip

We define a specific SM4 by fixing the \MS{} masses $\overline{m}_{t'}$, $\overline{m}_{b'}$, $\overline{m}_{\tau'}$, $\overline{m}_{\nu'}$ of the fourth generation fermions at the scale $M_Z$ where all the other SM parameters have previously been defined (see above). Next, we iterate over $\lambda$ from 0 to 8, and determine $m_H$ and $y_t(M_Z)$ from the Higgs and top matching corrections. After substituting $\mu=m_H$ in the first equation, \ref{eq:higgstopmatching} becomes a system of two equations in two unknowns, namely the scale $m_H$ where the first equation is satisfied, and the top Yukawa coupling $y_t(M_Z)$ whose evolution to $\mu=m_t$ must solve the second equation. We solve \ref{eq:higgstopmatching} numerically using a two-dimensional hybrid root-finding algorithm \cite{GSL} with $m_H=100$ GeV if $\lambda<2$, $m_H=1000$ GeV if $\lambda\geq2$, and $y_t(M_Z)=1$ as a first guess for the roots. Note that at this point in our analysis, we depart from the procedure described in Ref.~\cite{Hambye:1996wb} which solves \ref{eq:higgstopmatching} by an iterative application of a one-dimensional root-finding algorithm.

\medskip

The solution of the first equation in \ref{eq:higgstopmatching} gives us the physical Higgs mass $m_H$ that corresponds to $\lambda(M_Z)$, and the second equation fixes the top Yukawa coupling $y_t(M_Z)$. We now plug $\lambda(M_Z)$ and $y_t(M_Z)$ back into the RGEs and look at the evolution of $\lambda(\mu)$ with increasing scale $\mu$. At 2-loops, the quartic coupling does not have a Landau pole, but approaches an ultraviolet fixed point\footnote{The numerical value of \FP{} depends on the conventions in the RGEs.} $\FP{}=24.2944$. Perturbation theory breaks down long before $\lambda$ reaches the fixed point, and $\FP{}/2$ is expected to mark the beginning of the non-perturbative regime \cite{Riesselmann:1996is}. The smallest $\mu_0$ for which $\lambda(\mu_0)\geq\FP{}/2$ gives us the scale where the SM4 with a given $\lambda(M_Z)$ (and thus $m_H$) becomes non-perturbative. The set of all points $(\mu_0,m_H)$ constitutes the \textit{triviality bound}. Note that reading the graph the other way around gives an upper bound on the Higgs mass under the assumption that the theory be valid up to a given scale $\mu_0$.

\medskip

A lower bound on $m_H$ can be derived from the requirement of \textit{vacuum stability}. Although strictly speaking one should consider the effective Higgs potential and require that it does not become unbounded from below, the vacuum stability condition is well-approximated by the condition that $\lambda$ never becomes negative, where the agreement is better for larger scales and larger Higgs masses \cite{Casas:1996aq}. For obtaining the vacuum stability bound, we proceed exactly as before, except that now we record the set of points $(\mu_0,m_H)$ where $\mu_0$ is the smallest value for which $\lambda(\mu_0)<0$.

\section{Results}
\label{sec:results}

\subsection{The Standard Model}
\label{sec:results_SM3}

\begin{figure}[h!]
\centering
\includegraphics[width=.45\textwidth]{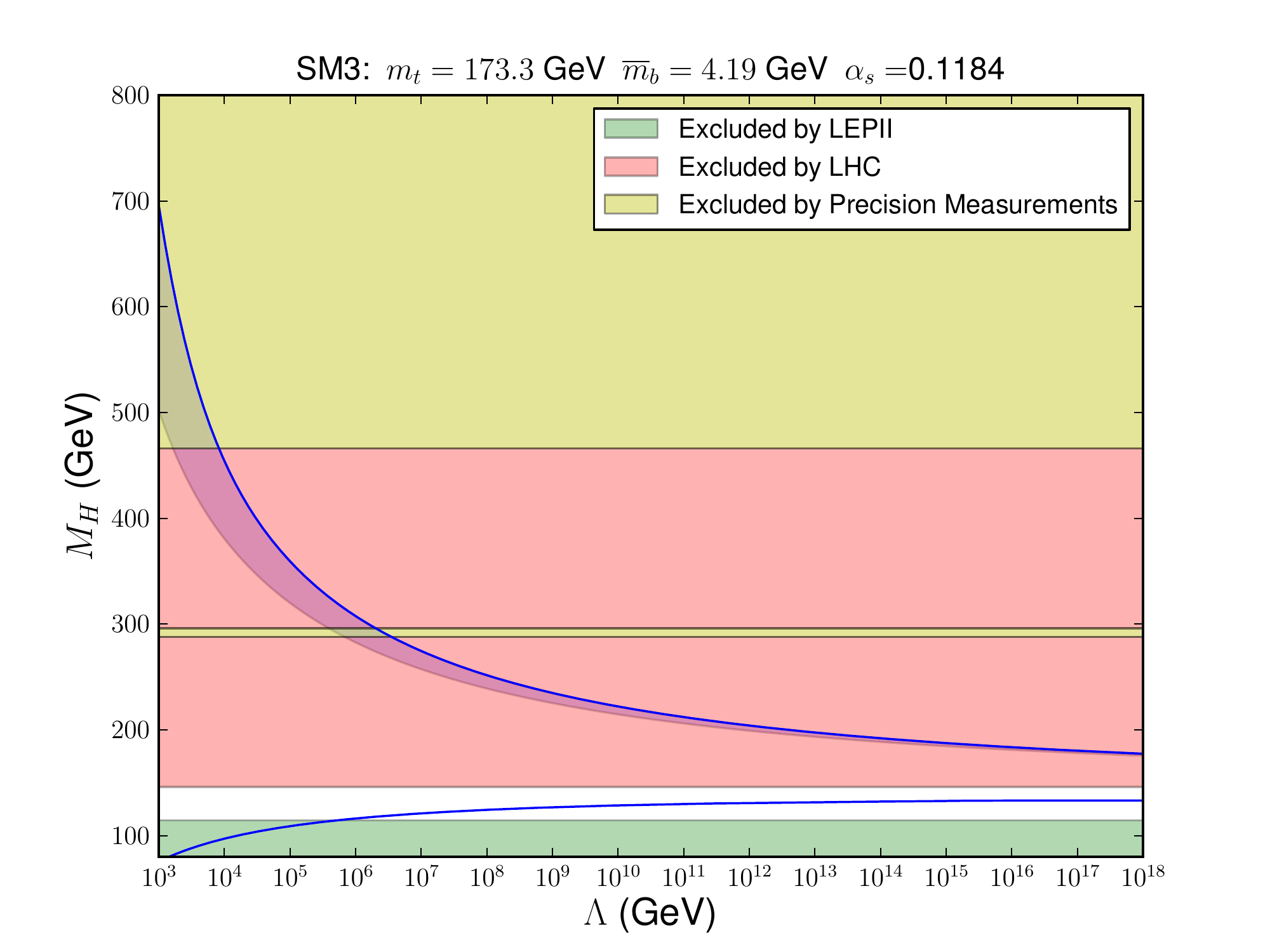}
\setcapindent{0em}
\caption{The stability and triviality bounds for the Standard Model. The horizontal bars indicate the Higgs exclusion limits from LEP \cite{Barate:2003sz}, LHC \cite{ATLAS-CONF-2011-135,CMS-TALK-LP2011}, and electroweak precision measurements \cite{aleph:2010vi}. The shaded area below the triviality bound indicates the uncertainty introduced by the choice of $\lambda < \FP/2$ or $\lambda < \FP/4$ as the criterion for perturbativity.
}
\label{fig:trivstab-SM3}
\end{figure}

We will first discuss the case of the Standard Model with three chiral generations of fermions before generalizing our results to the SM4. In \ref{fig:trivstab-SM3} we show the stability and triviality bounds that we obtained following the procedure described in \ref{sec:stability-and-triviality}. Varying the top mass $m_t=173.3\pm1.1$ GeV \cite{:1900yx} and the strong coupling constant $\alpha_s=0.1184\pm0.0007$ \cite{Nakamura:2010zzi} within their $1\sigma$ interval only has a small effect of $\leq2.6$ GeV and $\leq0.5$ GeV on the stability and triviality bound, respectively, and the corresponding error bands have therefore not been indicated in the plot. The theoretical uncertainty, however, is larger. As discussed before, $\lambda(\mu) \sim \FP/2$ is expected to be already very close to the non-perturbative regime, and applying the more conservative criterion $\lambda(\mu)<\FP/4$ leads to stricter bounds that we have indicated by the blue shaded area below the curve corresponding to the triviality bound. We will comment on the uncertainty of the stability curve at the end of this subsection.

\medskip

The Standard Model is most likely only an effective theory that is valid below a scale $\Lambda_c$ where new physics is expected to enter. From \ref{fig:trivstab-SM3} we can read off that for $\Lambda_c=1$ TeV, the Higgs mass must in the interval $76\leq m_H\leq696$ GeV, and analogously $97\leq m_H\leq454$ GeV for $\Lambda_c=10$ TeV, $109\leq m_H\leq359$ GeV for $\Lambda_c=100$ TeV, and finally $133\leq m_H\leq187$ GeV for $\Lambda_c=10^{15}$ GeV. Note, though, that we have obtained these limits under very specific assumptions like \textit{perturbative} validity of the SM up to $\Lambda_c$, \textit{absolute} stability of the vacuum, and the presence of only \textit{one} Higgs doublet in the theory. If one or more of these assumptions are relaxed, the bounds will be much weaker.

\medskip

The Higgs exclusion limits from LEP \cite{Barate:2003sz}, LHC \cite{ATLAS-CONF-2011-135,CMS-TALK-LP2011}, and electroweak precision measurements \cite{aleph:2010vi} leave only a narrow band of $114.4\leq m_H \leq 145$ GeV. These limits show that the triviality bound obtained from theory has now become obsolete and gives no further constraints. The stability bound intersects the upper edge of the LEP limit at $\Lambda_c=525$ TeV. For a Higgs with mass $m_H\geq133$ GeV there are no constraints from the stability bound either, and the Standard Model could at least in principle be valid up to some grand unification scale. However, should a Higgs boson lighter than $133$ GeV be discovered at LHC, then its mass will tell us the scale of new physics, e.g.~$m_H=130$ GeV corresponds to $\Lambda_c=1.2\times10^{11}$ GeV, $m_H=125$ GeV corresponds to $\Lambda_c=1.7\times10^{8}$ GeV, $m_H=120$ GeV corresponds to $\Lambda_c=5.9\times10^{6}$ GeV, and finally $m_H=115$ GeV corresponds to $\Lambda_c=6.5\times10^{5}$ GeV.

\medskip

We obtained the stability bound following the argument of Ref.~\cite{Altarelli:1994rb}, namely that to a good approximation the stability of the vacuum is equivalent to demanding that $\lambda(\mu)$ not become negative for any value of the scale $\mu$. As has been pointed out in Ref.~\cite{Casas:1996aq}, this simplified criterion corresponds to neglecting the 1-loop corrections to the 2-loop RGE improved effective potential. Including these 1-loop corrections \cite{Casas:1996aq} leads to a difference as large as 18 GeV at $\Lambda_c=1$ TeV, but this difference already decreases to 10 GeV at $\Lambda_c=10$ TeV and is well within the theoretical uncertainty band for $\Lambda_c\gtrsim100$ TeV. Since the stability bound provides a stronger limit than experiment only beyond $\Lambda_c>525$ TeV (see above), we can safely neglect this disagreement.

\medskip

To cross-check our results, we have reproduced the triviality bound with the same set of parameters and RGEs \cite{Machacek:1983tz,Machacek:1983fi,Machacek:1984zw,Ford:1992mv} as in Ref.~\cite{Hambye:1996wb}. In particular, we have not corrected for the typo pointed out in Ref.~\cite{Luo:2002ey} which was published after Ref.~\cite{Hambye:1996wb}. On comparing results, we find a small difference of a few GeV that we can trace back to the RGE evolution of the quartic coupling $\lambda(\mu)$. Fig.~(3a) of Ref.~\cite{Hambye:1996wb} shows $\lambda(M_Z)$ as a function of the top Yukawa coupling $g_t(M_Z)$. At 1-loop, we have perfect agreement. At 2-loop, we find a discrepancy between our result and the solid curve in Fig.~(3a) of Ref.~\cite{Hambye:1996wb} that starts out as zero at $g_t(M_Z)=0$, but grows with increasing $g_t(M_Z)$ and reaches a value of $\Delta\lambda(M_Z)=0.13$ (5.7\%) at $g_t(M_Z)=1.8$. Perusing the 2-loop RGE for $\lambda$ (see Eq.~(B.8) of Ref.~\cite{Machacek:1984zw}), we identify four terms (excluding the one proportional to $\lambda^3$) that can have an appreciable effect on the plot. Out of the $81$ possibilities of multiplying each of these terms by $0,\pm1$, there are 5 that bring about good agreement with the solid curve in Fig.~(3a) of Ref.~\cite{Hambye:1996wb}. In 2 of these 5 cases, the only modification to the RGE is \textit{one} change of sign.

\medskip

\subsection{The Standard Model with Four Chiral Generations}
\label{sec:results_SM4}

We now consider the case of the Standard Model with four chiral generations. Before we move on to discuss the results, we point out that all the fourth generation fermion masses are given in the \MS{} scheme at $\mu=M_Z$. The relation between the running and the pole mass is given e.g.~in Ref.~\cite{Melnikov:2000qh}. As a rule of thumb, the pole mass is 5-10\% larger than the running mass, and one needs to keep this difference in mind when comparing the numbers in the plots to experimental data.

\medskip

\begin{figure}[p!]
\centering
\subfigure[\footnotesize Stability and triviality bounds for SM4.]{
\includegraphics[width=.47\textwidth]{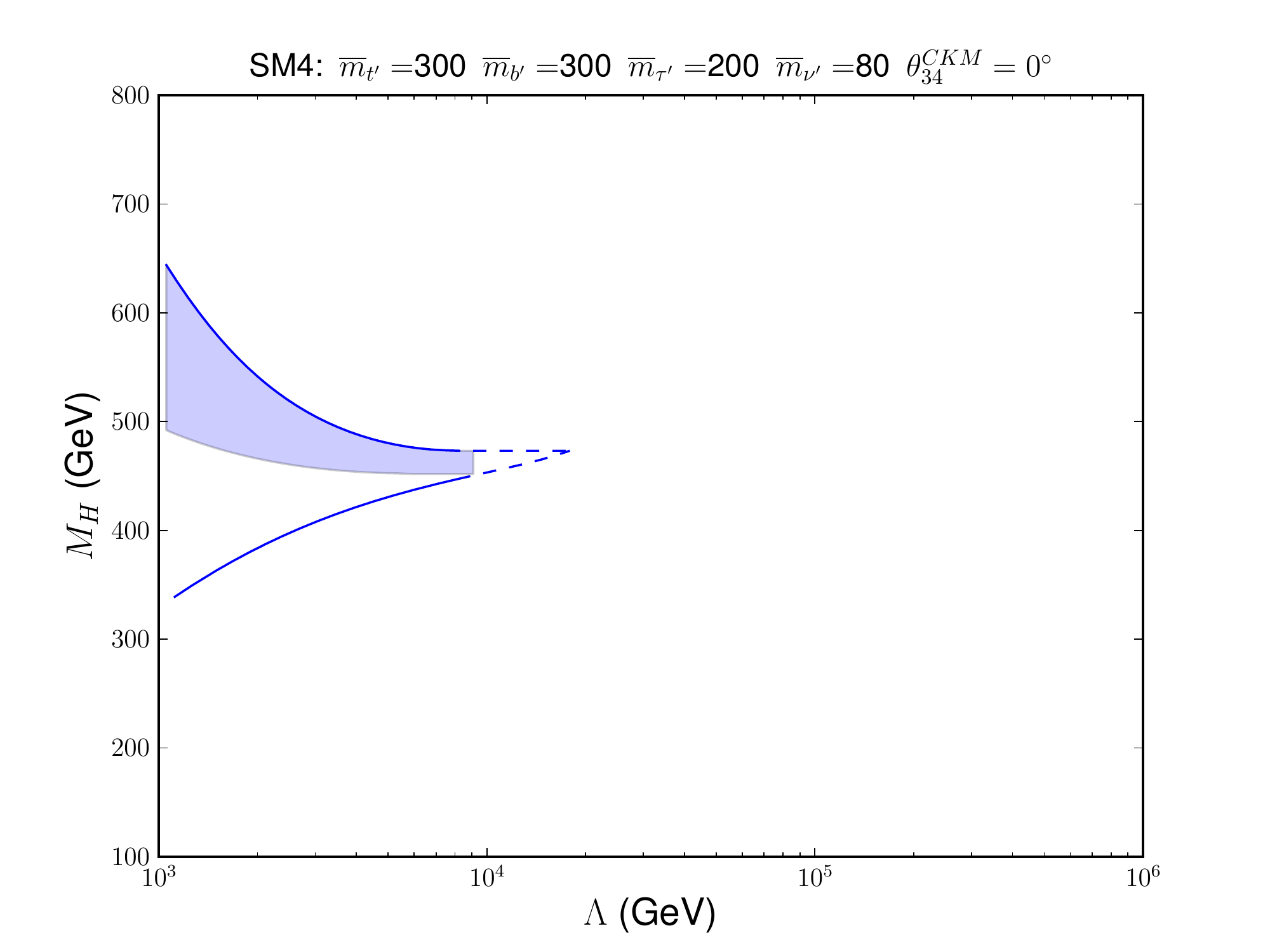}
\label{fig:trivstab-SM4-a}
}
\subfigure[\footnotesize Dependence on quark masses.]{
\includegraphics[width=.47\textwidth]{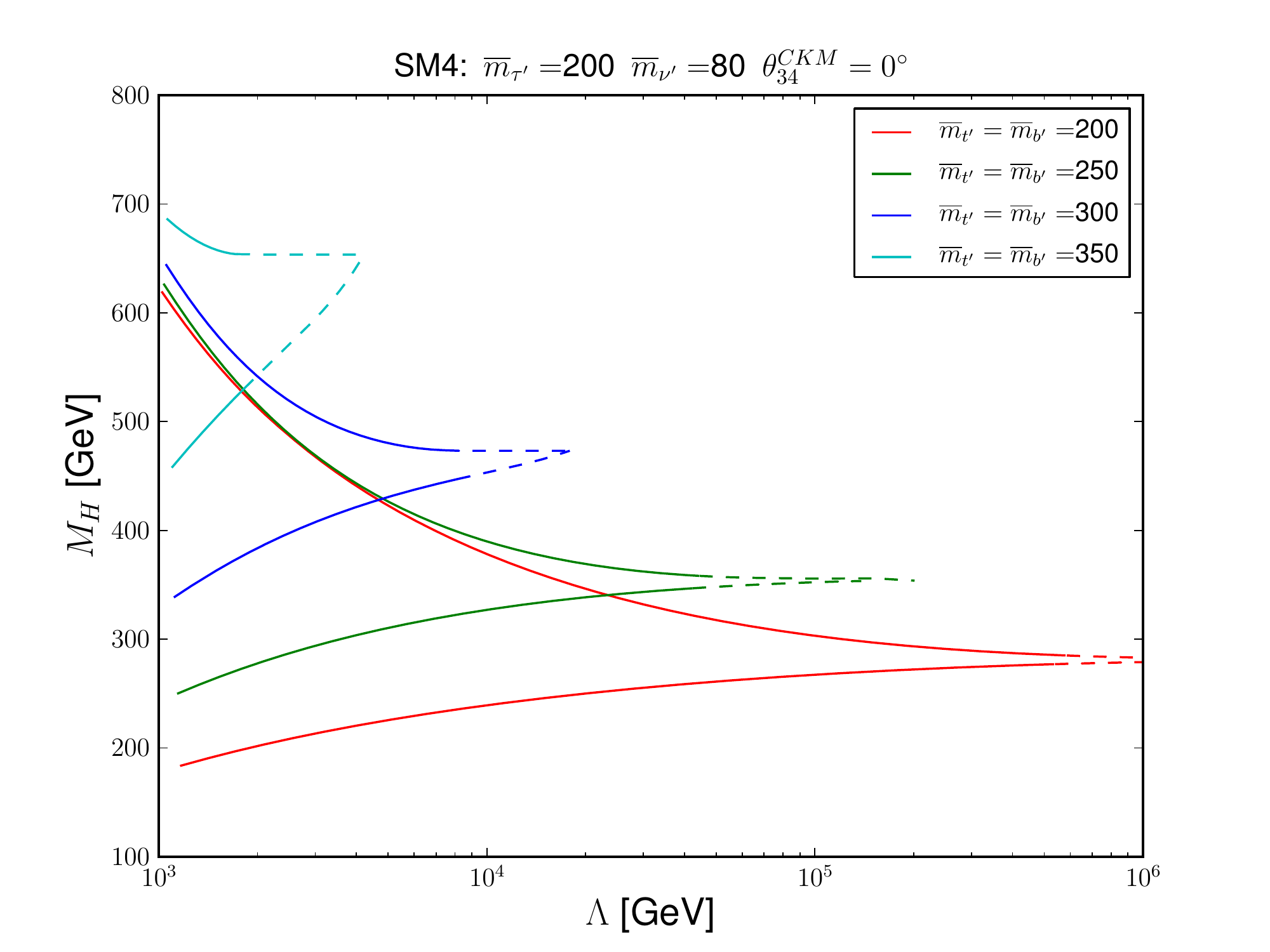}
\label{fig:trivstab-SM4-b}
}
\newline
\subfigure[\footnotesize Dependence on $b'$ mass.]{
\includegraphics[width=.47\textwidth]{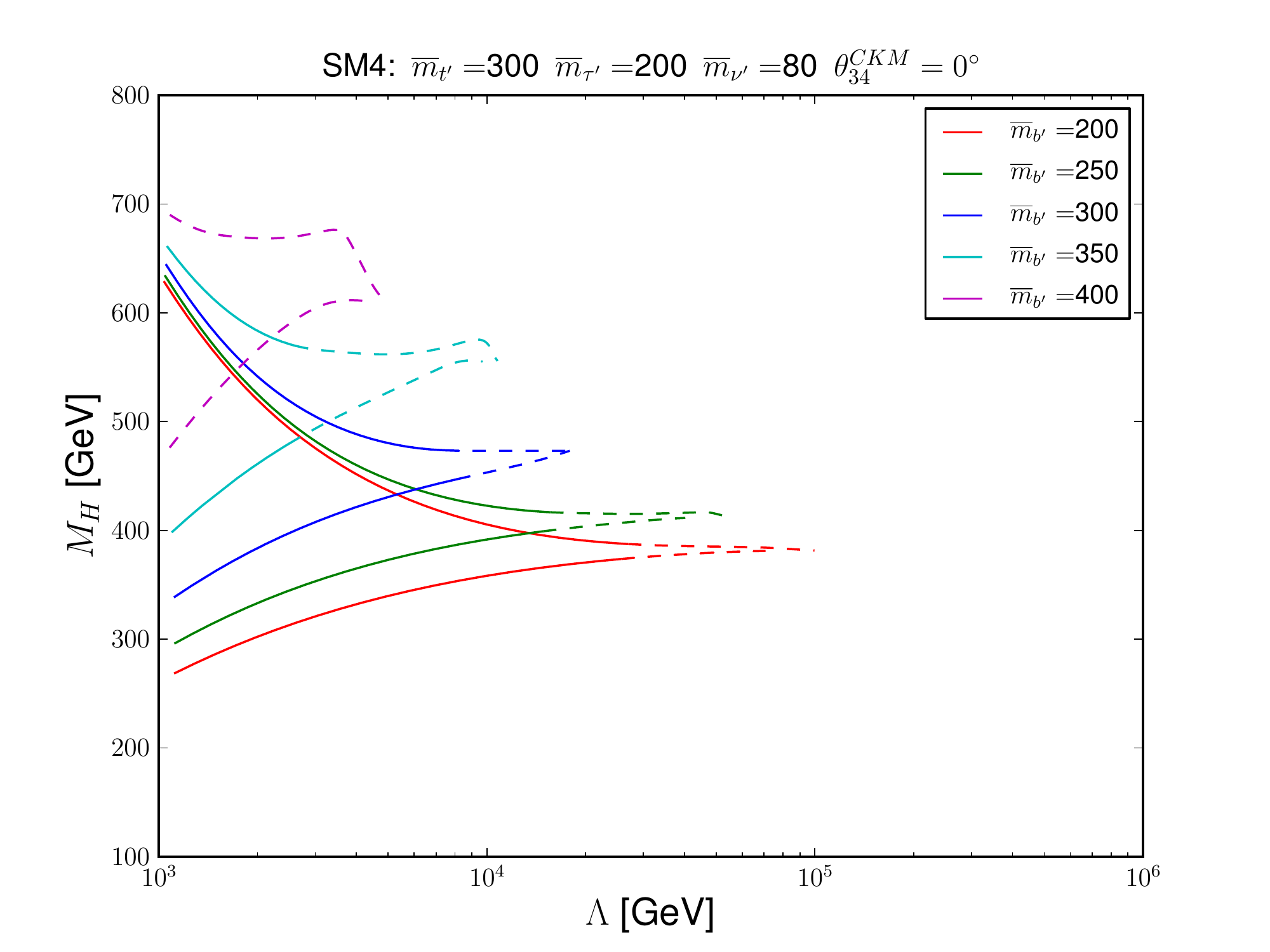}
\label{fig:trivstab-SM4-c}
}
\subfigure[\footnotesize Dependence on $\tau'$ mass.]{
\includegraphics[width=.47\textwidth]{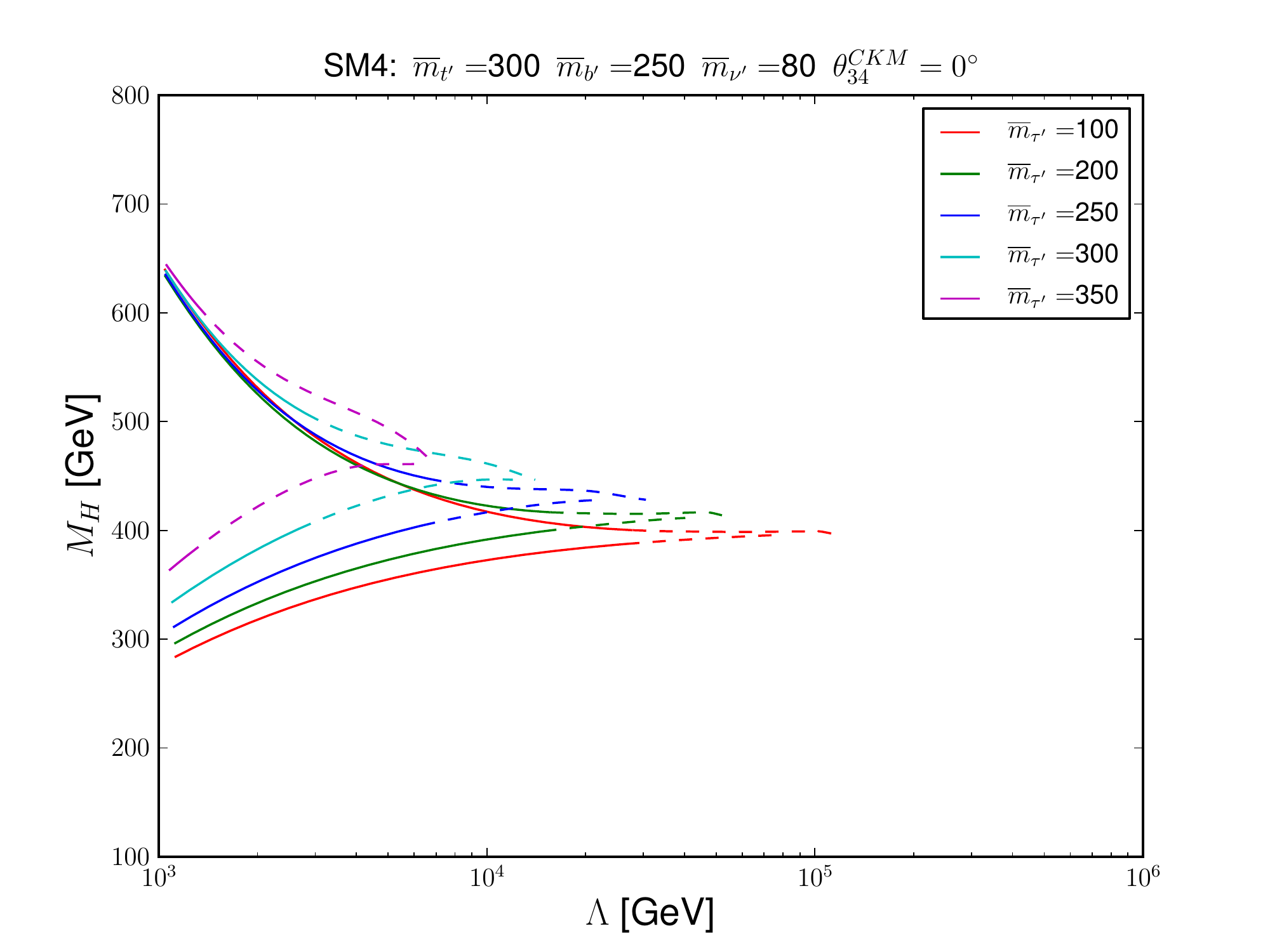}
\label{fig:trivstab-SM4-d}
}
\newline
\subfigure[\footnotesize Dependence on $\nu'$ mass.]{
\includegraphics[width=.47\textwidth]{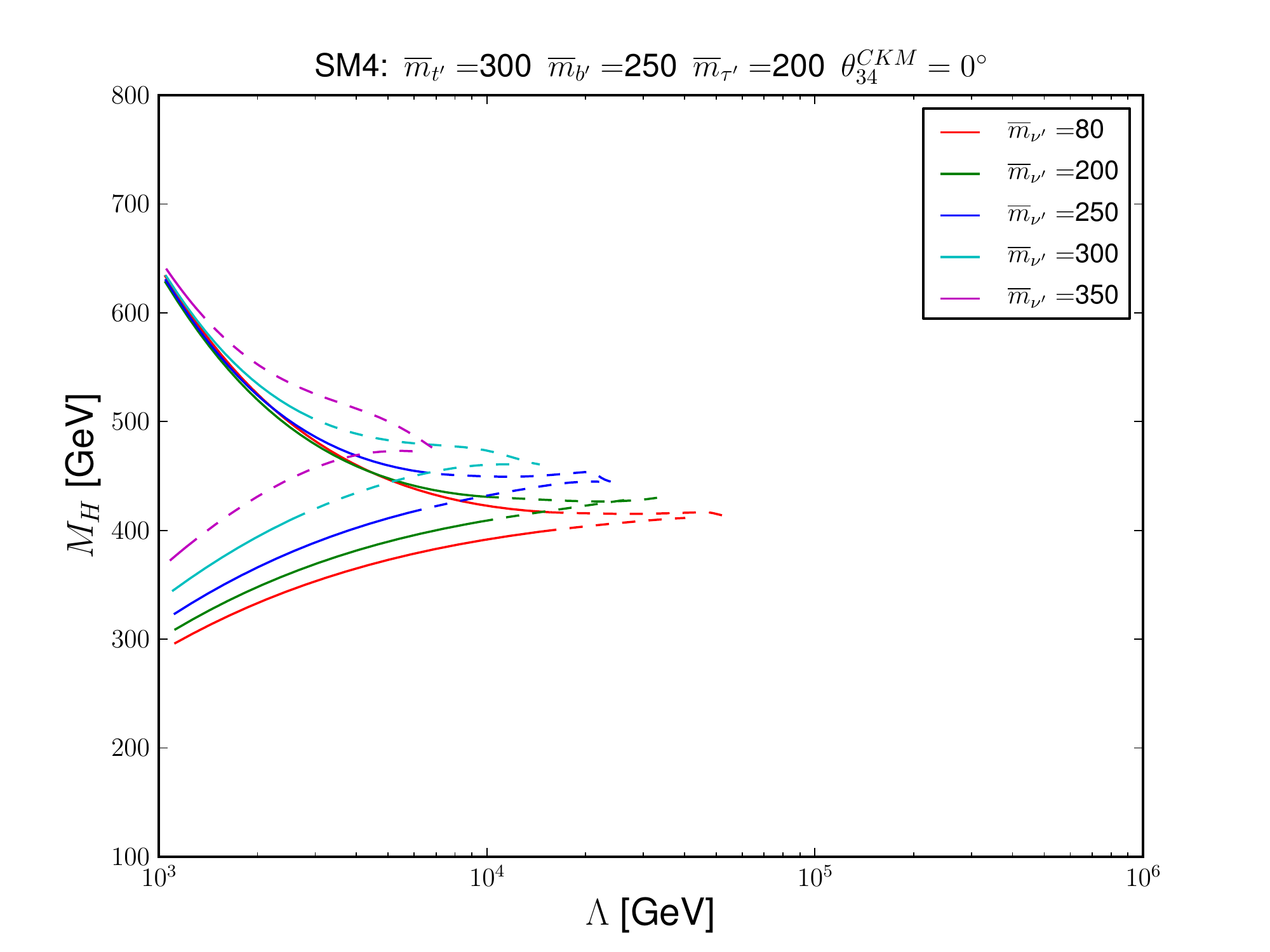}
\label{fig:trivstab-SM4-e}
}
\subfigure[\footnotesize Dependence on quark mixing.]{
\includegraphics[width=.47\textwidth]{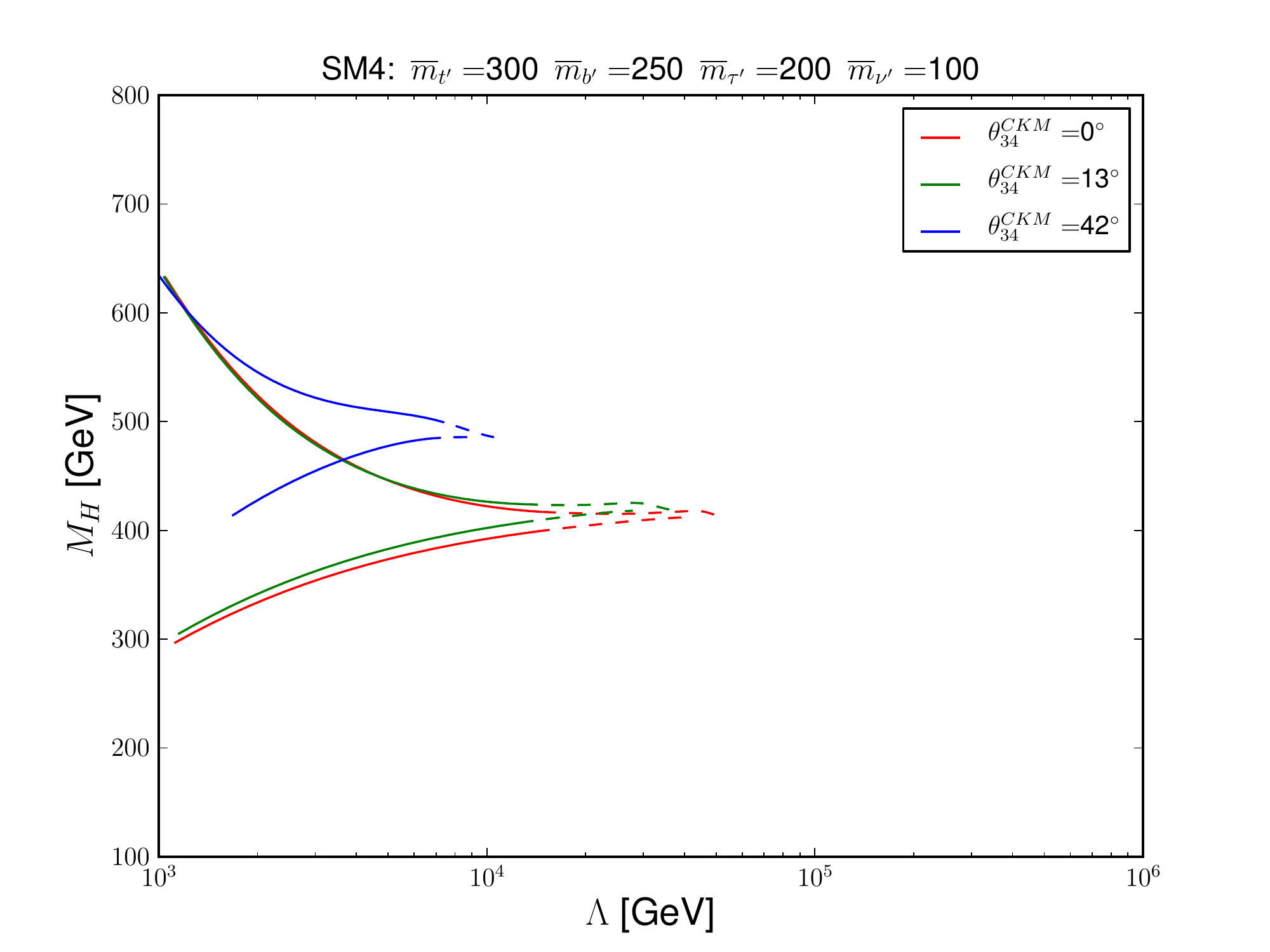}
\label{fig:trivstab-SM4-f}
}
\setcapindent{0em}
\caption{Stability and triviality bounds for the SM4. The dashed lines indicate where at least one of the Yukawa couplings becomes non-perturbative (i.e.~$Y_{f}\geq\sqrt{4\pi}$ for $f=t,t',b',\tau',\nu'$). All fermion masses are given in the \MS{} scheme (in units of GeV). More plots can be viewed online \cite{aw:2011:sm4plots}.
}
\label{fig:trivstab-SM4}
\end{figure}

In \ref{fig:trivstab-SM4-a} we show the stability and triviality bounds for a specific choice of fourth generation fermion masses. The shaded band below the triviality bound indicates its theoretical uncertainty. The choice of lepton masses satisfies both the limits from direct searches (see \vref{eq:explimitsfermions}) and the constraints from electroweak precision measurements for $m_{t'}=m_{b'}$ (see Fig.~(13) of Ref.~\cite{Baak:2011ze}). For the quark masses, we have taken a value of 300 GeV which is excluded except maybe in a small corner of the parameter space (see detailed discussion in \ref{sec:directlimits}). However, for larger quark masses the theory quickly becomes strongly-coupled at moderately low energies so that the stability and triviality bounds loose their meaning. In \ref{fig:trivstab-SM4-a} we have indicated the energy scales where any of the Yukawa couplings becomes non-perturbative (i.e.~$Y_{f}\geq\sqrt{4\pi}$ for $f=t,t',b',\tau',\nu'$) by changing from a solid to a dashed line-style. 

\medskip

The most striking feature in \ref{fig:trivstab-SM4-a} is that now the two curves intersect and bound a finite region. Strictly speaking we cannot trust the stability and triviality bounds beyond $\Lambda_c\simeq9$ TeV where they become dashed; the best we can say is that if no new physics enters before that scale, the Higgs mass should lie between 438 and 456 GeV. It is tempting to assume the validity of the bounds until they meet at $\Lambda_c\simeq25$ TeV, but we have checked that relaxing the criterion for the perturbativity of the Yukawa couplings from $Y_{f}\geq\sqrt{4\pi}$ to $Y_{f}\geq4\pi$ does not make a noticeable difference in the results.

\medskip

In \ref{fig:trivstab-SM4-b} we show how the stability and triviality bounds change when we fix the lepton masses at $m_{\tau'}$=200 GeV and $m_{\nu'}=80$ GeV, and vary the quark masses $m_{t'}=m_{b'}$ from 200 to 350 GeV. We see that the stability bound depends more sensitively on the quark masses than the triviality bound. Not surprisingly, with increasing quark masses the lower limit on the Higgs mass also increases, and at the same time the scale where the perturbativity of the Yukawa couplings breaks down gets lower and lower, until it reaches $\Lambda_c\simeq2$ TeV for $m_{t'}=m_{b'}=350$ GeV. For quark masses at their experimental lower limits, perturbativity is lost long before.

\medskip

\ref{fig:trivstab-SM4-c} shows the dependence of the stability and triviality bounds on the $b'$ mass while holding the other parameters fixed. The lepton masses $m_{\tau'}$, $m_{\nu'}$ will not be compatible with the electroweak precision measurements for all choices of $m_{b'}$ (cf.~Fig.~(13) of Ref.~\cite{Baak:2011ze}), but for the present purpose of showing the variation of the curves with the $b'$ mass, this is acceptable. We find that the effect is comparable in magnitude to the previous case considered in \ref{fig:trivstab-SM4-b}.

\medskip

As we can see in \ref{fig:trivstab-SM4-d}, varying $m_{\tau'}$ has a smaller effect on the stability and triviality bounds than the quarks masses. For completeness we have also included neutrino masses in the RGE running, since in the SM4 $m_{\nu'}$ is necessarily large and cannot be neglected. \ref{fig:trivstab-SM4-e} shows that $\nu'$ will have a noticeable effect for $m_{\nu'}\gtrsim200$ GeV.

\medskip

In \ref{fig:trivstab-SM4-f} we show the stability and triviality bounds for three different mixing scenarios, namely no mixing, small mixing $\sim13^\circ$ \cite{Chanowitz:2009mz} and large mixing $\sim42^\circ$ \cite{Bobrowski:2009ng}. A large mixing angle $\theta_{34}^{\textrm{CKM}}=42^\circ$ is ruled out by electroweak precision measurements \cite{Chanowitz:2009mz}, and the difference between the no mixing and small mixing scenarios is small. In particular, the very small mixing angles considered in Ref.~\cite{Hung:2007ak} will not affect the curves at all.

\begin{figure}[h!]
\centering
\includegraphics[width=1.\textwidth, trim=40mm 0mm 45mm 0mm]{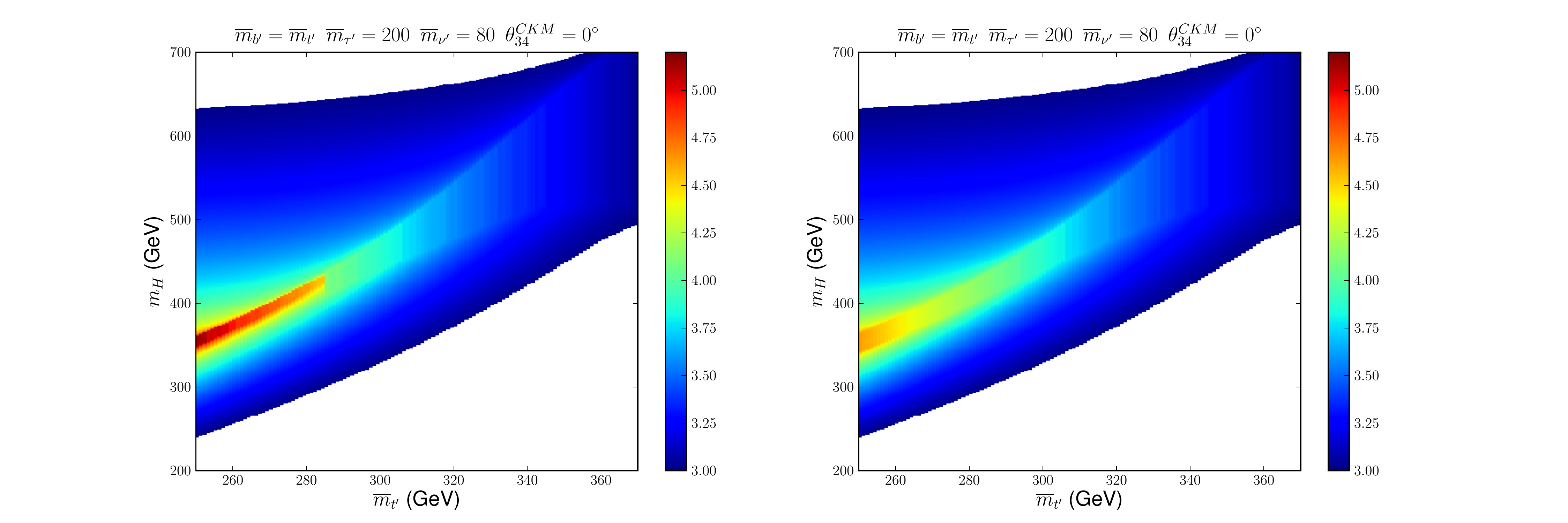}
\setcapindent{0em}
\caption{The left panel shows the maximal scale $\Lambda_\textrm{max}$ for which the Higgs mass $m_H$ for a given $\overline{m}_{t'}=\overline{m}_{b'}$ still lies inside the area bounded by the triviality and stability bounds. The other SM4 parameters are kept fixed: $\overline{m}_{\tau'}=200$ GeV, $\overline{m}_{\nu'}=80$ GeV, $\theta_{34}^{\textrm{CKM}}=0^\circ$. In the right panel, we present the same maximal scale $\Lambda_\textrm{max}$ where we have now additionally required that all Yukawa couplings stay in the perturbative regime. The bar to the right of each plot gives the correspondence between the colors and the logarithm to base 10 of $\Lambda_\textrm{max}$ [GeV].}
\label{fig:mHvsmt4}
\end{figure}

\medskip

With increasing fourth generation fermion masses the stability and triviality bounds move towards higher Higgs masses. At the same time the scale where the perturbativity is lost rapidly decreases. In \ref{fig:mHvsmt4} we show for fixed but arbitrary $m_H$, $\overline{m}_{t'}$, $\overline{m}_{b'}$ the maximal scale $\Lambda_\textrm{max}$ where the SM4 ceases to be meaningful and new physics must necessarily enter. In the left panel we do not demand that the Yukawa couplings be perturbative whereas in the right panel we do. In other words, for given masses $\overline{m}_{t'}$ and $\overline{m}_{b'}$ we record the scale $\Lambda_\textrm{max}$ where the line corresponding to a constant Higgs mass $m_H$ intersects the stability or triviality curve. The white area in \ref{fig:mHvsmt4} corresponds to values of $m_H$, $\overline{m}_{t'}$, $\overline{m}_{b'}$ that lie outside the finite area bounded by the stability and triviality curves. From the plots we can read off that if we want the SM4 to be valid up to 10 TeV, the quark masses cannot be much higher than $\sim300$ GeV. Also, we see that $\overline{m}_{t'}\sim\overline{m}_{b'}\sim375$ GeV is as high as we can go before perturbativity is lost before 1 TeV. We have checked that varying the lepton masses $\overline{m}_{\tau'}$, $\overline{m}_{\nu'}$ does not change this conclusion.

\medskip

Note that from \ref{fig:trivstab-SM4} and \ref{fig:mHvsmt4} we also learn that the Higgs mass cannot exceed 700 GeV for any values of the fourth generation fermion masses, if the Yukawa couplings are to remain perturbative.

\medskip

The stability and triviality bounds in the context of a fourth generation have been considered in the literature before \cite{Kribs:2007nz}. Our work improves upon the previous results in the following ways:
\begin{inparaenum}[(i)]
\item We use the full 2-loop RGEs to run all SM parameters.
\item We use 1-loop matching corrections for the Higgs and the top quark.
\item We include the effect of the massive fourth generation neutrino.
\item We consider non-trivial mixing between the third and the fourth generation.
\item We indicate the constraints stemming from the perturbativity of the Yukawa couplings.
\end{inparaenum}
The numerical differences can be significant, as a comparison with Ref.~\cite{Kribs:2007nz} shows. For  $\overline{m}_{t'}=\overline{m}_{b'}=260$ GeV, from Fig.~(7) of Ref.~\cite{Kribs:2007nz} we read off $200\leq m_H\leq470$ GeV at 1 TeV and $\Lambda_c=300$ TeV. In our case, we obtain from \ref{fig:trivstab-SM4-b} the values $250\leq m_H\leq620$ GeV at 1 TeV and $\Lambda_c=200$ TeV. Furthermore, the perturbativity of the Yukawa couplings is already lost at $\Lambda_c=40$ TeV and the stability and triviality bounds should not be trusted beyond that point.

\subsection{Theoretical Constraints on Fourth Generation Fermion Masses}
\label{sec:theolimits}

In \ref{sec:directlimits} we discussed in detail the direct limits on the fourth generation fermion masses set by Tevatron and LHC and how they may not be applicable in some corners of the parameter space where the mixing between the third and fourth generation is very small. Recent analyses at LHC \cite{ATLAS-CONF-2011-135,CMS-HIG-11-011} exclude a SM4 Higgs boson in the mass range $120\leq m_H\leq 600$ GeV. From \ref{fig:trivstab-SM4-b} and \ref{fig:trivstab-SM4-c} we see that that if we assume the validity of the SM4 up to a scale of 2 TeV, the stability and triviality bounds for quark masses smaller than 300 GeV are not compatible with the Higgs exclusion limits from experiment. Analogously, from \ref{fig:trivstab-SM4-d} and \ref{fig:trivstab-SM4-e} we read off that the leptons $\tau'$ and $\nu'$ should be heavier than 350 GeV. These lower limits refer to the \MS{} masses, and the corresponding limits on the pole masses are by 5-10\% higher. An upper limit can be obtained by noting that that perturbativity is lost before $1$ TeV for quark and lepton masses as low as 400 GeV.

\medskip

The limits on the quark masses are competitive with the direct limits obtained at Tevatron, and those on the lepton masses are better than any collider limits (cf.~\vref{eq:explimitsfermions}). Furthermore, these limits are applicable for all values of the mixing angles including the small mixing scenario.

\medskip

To set more stringent limits, we need Higgs searches that go beyond 600 GeV. If no Higgs with $m_H\leq700$ GeV is observed, a fourth generation is ruled out for any values of the fermion masses, provided that the Yukawa couplings are in the perturbative regime.

\section{Conclusions}
\label{sec:conclusions}

The recent Higgs results from LHC together with mild theoretical assumptions allow us to exclude a fourth generation with quark masses $\overline{m}_{t'},\overline{m}_{b'}\leq300$ GeV and lepton masses $\overline{m}_{\tau'},\overline{m}_{\nu'}\leq350$ GeV. Our analysis covers regions of parameter space where the experimental limits are not applicable. Furthermore, the bounds we obtain for the lepton masses are stronger than the direct limits from experiment. The direct limits from fourth generation quark searches at LHC are already closing in on the unitarity bound and will in near future rule out most (but not all) of the parameter space of a fourth generation. In order to conclusively rule out a fourth generation with perturbative Yukawa couplings the Higgs analyses at LHC need to be extended to cover the region $m_H\leq700$ GeV. In the context of the Standard Model with three generations, the discovery of the Higgs with a mass below 133 GeV will indicate the scale where new physics must necessarily enter.

\bigskip
\bigskip
\textbf{Acknowledgments} \\

\noindent
I am indebted to Rohini M.~Godbole for suggesting the research topic, and to Sabine Kraml and Sudhir K.~Vempati for collaboration in the initial stages of this project. I acknowledge useful discussions with Olivier Bondu, Athanasios Dedes, Tom\'a\v{s} Je\v{z}o, J\"org Meyer, Markus Schumacher and Pietro Slavich. I thank Thomas Hambye, Yury F.~Pirogov and Oleg Zenin for correspondence. I thank the \emph{Centre de Calcul de l'Institut National de Physique Nucl\'{e}aire et Physique des Particules} in Lyon for using their resources.


\appendix

\labelformat{section}{Appendix #1} 

\section{Renormalization Group Equations}
\label{sec:RGESM4}

The 2-loop renormalization group equations for the SM4 can be easily obtained from the standard references \cite{Machacek:1983tz,Machacek:1983fi,Machacek:1984zw} by substituting $n_G=4$ and promoting the Yukawa matrices to 4-by-4 matrices. In the SM4, the fourth generation neutrino is necessarily heavy, and its effect on the RGEs cannot be neglected. We have used the 2-loop RGEs for the neutrinos as given by Ref.~\cite{Pirogov:1998tj}.

\medskip

In our analysis, we followed Refs.~\cite{Machacek:1983tz,Machacek:1983fi,Machacek:1984zw,Pirogov:1998tj} \textit{except} for a number of typos that we will list in the following. Note that most of these typos have been reported before and some of them may be self-evident, but we think that it is important to include this information in order to unambiguously specify the RGEs we used.

\medskip

In Ref.~\cite{Machacek:1983tz}, the coefficients $C^U_k$, $C^D_k$, $C^L_k$ in Eq.~(B.2) should read  $C^U_\ell$, $C^D_\ell$, $C^L_\ell$, respectively. Above Eq.~(B.3a), the relation $g_1=\sqrt{3/5}\,g'_{W-S}$ should read $g_1=\sqrt{5/3}\,g'_{W-S}$.

\medskip

In Ref.~\cite{Machacek:1983fi}, the term $-2\lambda(3H^\dagger H + F_D^\dagger F_D)$ in Eq.~(B.8) should read $-6\lambda H^\dagger H$ \cite{Luo:2002ey}. The term $-2\lambda(3F_D^\dagger F_D + H^\dagger H)$ in Eq.~(B.9) should read $-6\lambda F_D^\dagger F_D$ \cite{Luo:2002ey}.

\medskip

In Ref.~\cite{Machacek:1984zw}, the term $(9/5\,g_1^2+g_2^2)\lambda$ in Eq.~(B.3) should read $(9/5\,g_1^2+9\,g_2^2)\lambda$ \cite{Ford:1992mv}. In Eq.~(B.4), the term $3F_D F_D$ should read $3F_D^\dagger F_D$. In Eq.~(B.8), the term $(229/24+2n_G)g'^4$ should read $(229/24+50/9\,n_G)g'^4$. According to Ref.~\cite{Arason:1991ic}, the latter term should read $(229/4+50/9\,n_G)g'^4$, but Ref.~\cite{Ford:1992mv} lists the same term for the special case $n_G=3$ as $629/24\,g'^4\,\lambda$ which disagrees with Ref.~\cite{Arason:1991ic}. In Eq.~(B.8) there are two more terms, namely $39/4\,g^2 g'^2$ and $3/2\,g^4 Y_2(S)$, which disagree with Ref.~\cite{Arason:1991ic} but do agree with Refs.~\cite{Ford:1992mv,Luo:2002ey}. We have followed Ref.~\cite{Luo:2002ey}. Finally, the term $6\lambda\mathrm{Tr}(H^\dagger H F_D^\dagger F_D)$ in Eq.~(B.8) should read $-42\lambda\mathrm{Tr}(H^\dagger H F_D^\dagger F_D)$ \cite{Luo:2002ey}. 

\medskip

In Appendix A.1 of Ref.~\cite{Pirogov:1998tj}, the term $-9\sum ( 3y_{u_g}^4 + 3y_{u_g}^4 )$ in the 2-loop RGE for the $\tau$ Yukawa coupling should read $-9\sum ( 3y_{u_g}^4 + 3y_{d_g}^4 )$. In Appendix A.2, in the 2-loop RGEs for the Yukawa couplings of the charged leptons, the terms that derive from the generalizations of $Y_4(S)$, $H(S)$, $\chi_4(S)$ \cite{Machacek:1983tz,Machacek:1983fi,Machacek:1984zw} to include the neutrino Yukawa matrices are missing, whereas the ones deriving from $Y_2(S)$ are present. An example is the term $(3/8\,g^2_1+15/8\,g^2_2)\textrm{Tr}(Y_\nu^\dagger Y_\nu)$ that is introduced by $Y_4(S)$. In the same equation, there is a likely misprint\footnote{We thank Oleg Zenin for pointing this out.} and the term $(-3/16\,g_1^2+129/16\,g_2^2)Y_\nu^\dagger Y_\nu$ should read $(-3/16\,g_1^2+9/16\,g_2^2)Y_\nu^\dagger Y_\nu$.

\section{Generalization of the Higgs Matching Correction}
\label{sec:higgsmatchingconditions}

Beyond leading order, the relation between the physical Higgs mass $m_H$ and the quartic self-coupling $\lambda$ receives radiative corrections that can be neatly expressed as \cite{Sirlin:1985ux}
\begin{equation}
\lambda(\mu) = \frac{m_H^2}{v^2} \left(1+\delta_H(\mu) \right).
\end{equation}
For the case of the SM3, $\delta_H$ is explicitly given by Eqs.~(15a)-(15f) of Ref.~\cite{Sirlin:1985ux}. Unfortunately, these formulas do not directly generalize to the case of the SM4 so that we have to go back to the original expression for $\delta_H$ that is given in Eq.~(14b) of Ref.~\cite{Sirlin:1985ux}:
\begin{equation}
\delta_H(M) = \left. -\frac{1}{m_H^2}\textrm{Re}\left[\Pi_{HH}(m_H^2)\right] - \frac{1}{m_H^2} \frac{T}{v} + \frac{1}{m_W^2} \left( A^{\textrm{bos}}_{WW}(0) + A^{\textrm{had}}_{WW}(0)  + A^{\textrm{lep}}_{WW}(0) \right) - E  \enspace \right|_\textrm{finite}
\end{equation}

\medskip

Expressions for $T$ and $\text{Re}\left[\Pi_{HH}(m_H^2)\right]$ are given in Eq.~(A.1) and Eq.~(A.2) of Ref.~\cite{Sirlin:1985ux}, respectively, and are easily generalized to the case of the SM4 by extending the sum over the fermions to include the fourth generation particles. The expression for $E$ is given in Eq.~(9d) of Ref.~\cite{Sirlin:1985ux} and remains unchanged.

\medskip

$A^\text{bos}_{WW}(0)$ is given in Eq.~(A13) of Ref.~\cite{Marciano:1980pb} and remains unchanged in the presence of an extra generation of fermions.

\medskip

$A^\text{had}_{WW}(0)$ is given in Eq.~(B2) in ref.~\cite{Marciano:1980pb} as the coefficient of $g_{\mu\nu}$ for $q^2=0$. In the case of the SM3, all quark masses except for the top mass can be neglected, and the formulas simplify accordingly. For the SM4, we have to include the additional massive quarks, and we cannot neglect the bottom mass, since the fourth generation quarks may have non-trivial mixing with the third generation:
\begin{multline}
\textstyle
A^\text{had}_{WW}(0) = -\frac{3 g^2}{16 \pi ^2} \left( \frac{V_{bt}^2 \left(-2 m_b^4 \log \left(m_b^2/M^2\right)+2 m_t^4 \log
   \left(m_t^2/M^2\right)+m_b^4-m_t^4\right)}{4 \left(m_t^2-m_b^2\right)}\right.\\ 
\textstyle
   \left. + \frac{V_{bt'}^2 \left(-2 m_b^4 \log
   \left(m_b^2/M^2\right)+2 m_{t'}^4 \log \left(m_{t'}^2/M^2\right)+m_b^4-m_{t'}^4\right)}{4
   \left(m_{t'}^2-m_b^2\right)}+\frac{V_{b't}^2 \left(-2 m_{b'}^4 \log \left(m_{b'}^2/M^2\right)+2 m_t^4 \log
   \left(m_t^2/M^2\right)+m_{b'}^4-m_t^4\right)}{4 \left(m_t^2-m_{b'}^2\right)}\right.\\
\textstyle
\left.+\frac{V_{b't'}^2 \left(-2 m_{b'}^4 \log
   \left(m_{b'}^2/M^2\right)+2 m_{t'}^4 \log \left(m_{t'}^2/M^2\right)+m_{b'}^4-m_{t'}^4\right)}{4
   \left(m_{t'}^2-m_{b'}^2\right)}\right)
\label{eq:AWWhad}
\end{multline}

\medskip

Here, $V_\textrm{CKM}$ (or $V$ for short) denotes the quark mixing matrix. $A^\text{lep}_{WW}(0)$ is easily obtained from \ref{eq:AWWhad} by performing the evident substitutions $V_\textrm{CKM}\rightarrow U_\textrm{PMNS}$, $m_{t'}\rightarrow m_{\nu'}$, $m_{b'}\rightarrow m_{\tau'}$, neglecting the third generation lepton masses and canceling a factor of 3 that corresponds to the color degrees of freedom.

\medskip

The resulting full expression for $\delta_H$ in the case of the SM4 is quite lengthy and has therefore been relegated to a separate document that has been made available for download \cite{aw:2011:smrges}.

\begin{figure}[h!]
\centering
\subfigure[\footnotesize Higgs matching corrections $\delta_H(\mu)$ for SM3.]{
\includegraphics[width=.45\textwidth]{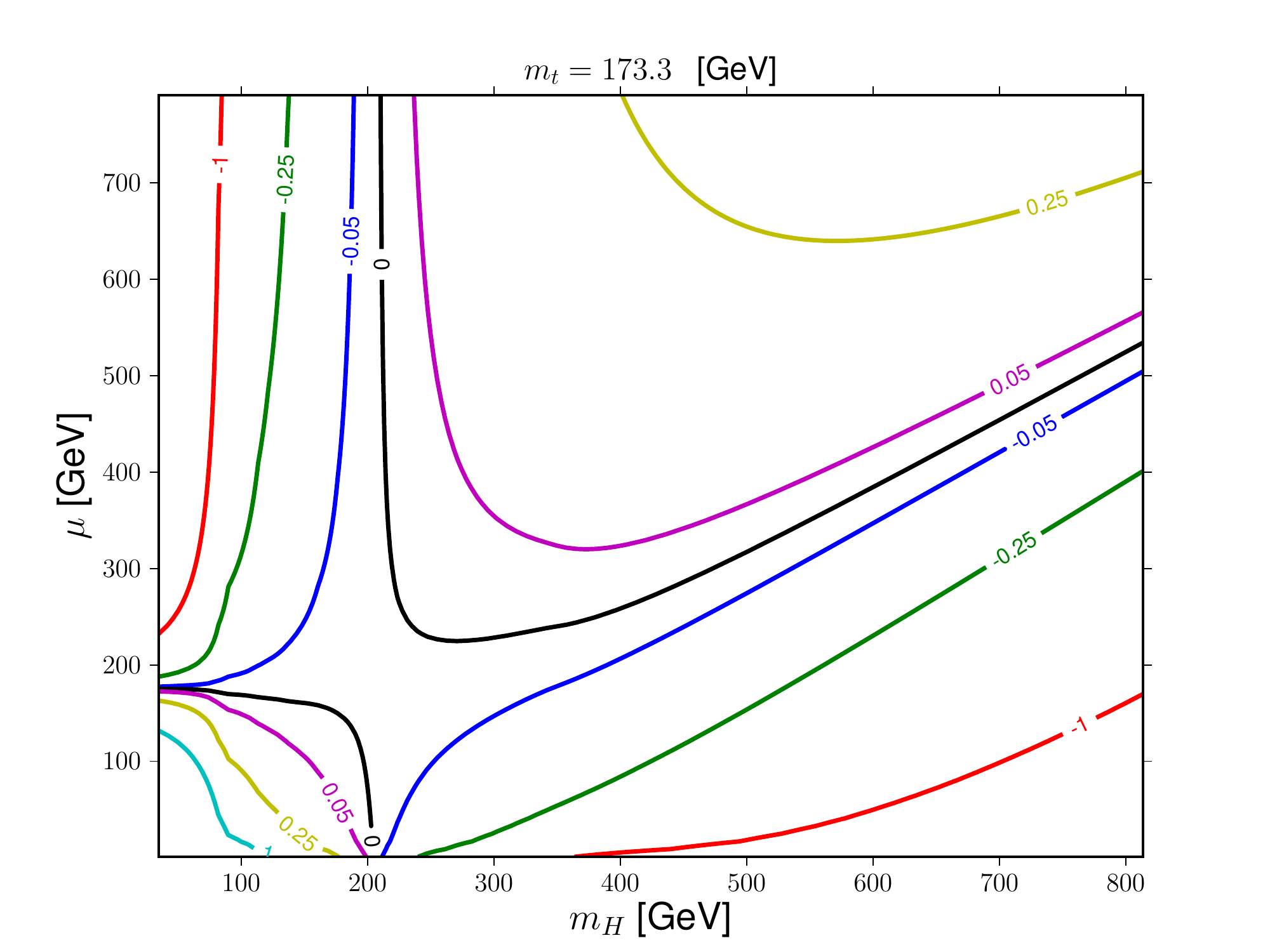}
\label{fig:2a}
}
\subfigure[\footnotesize Higgs matching corrections $\delta_H(\mu)$ for SM4.]{
\includegraphics[width=.45\textwidth]{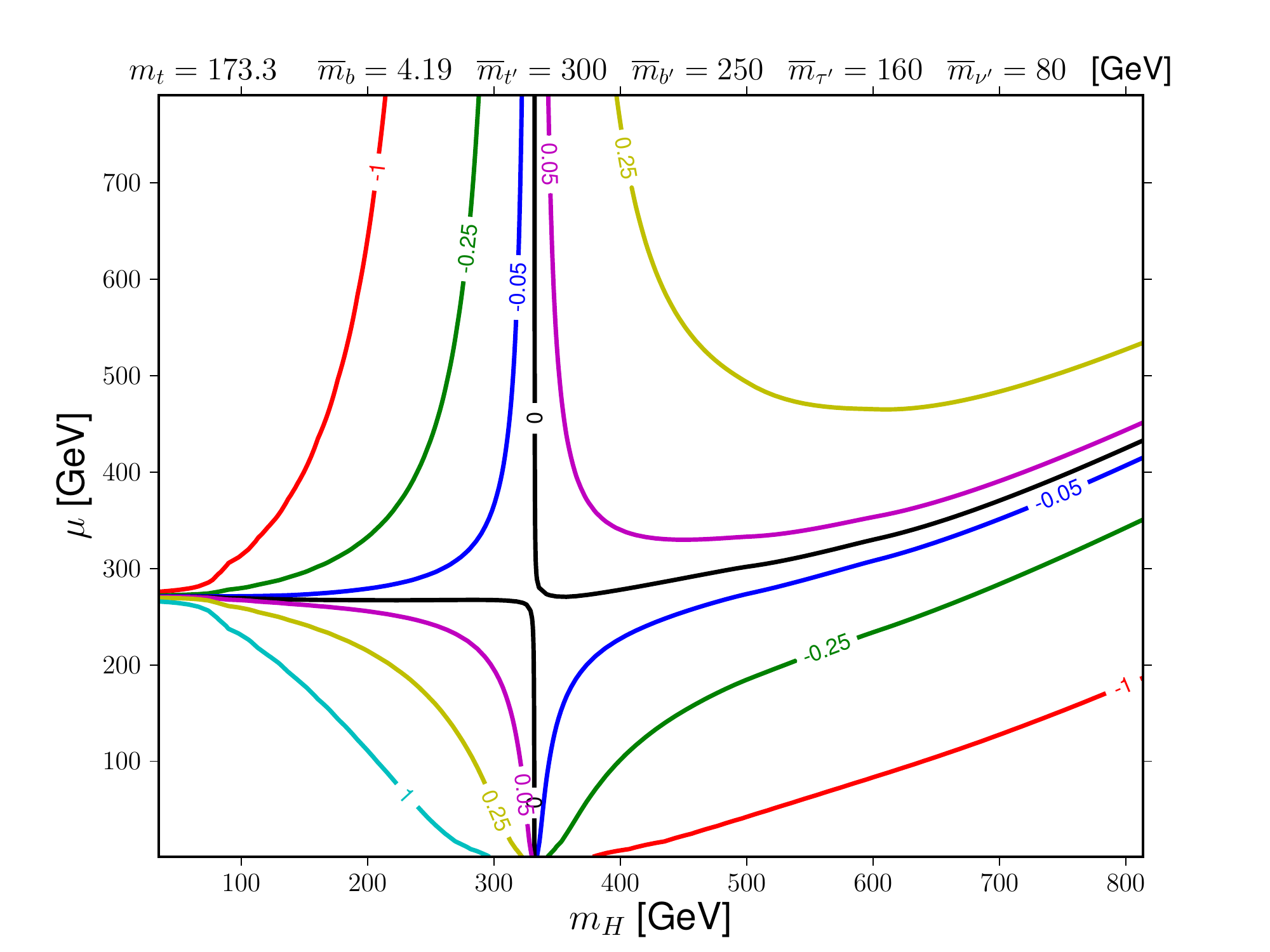}
\label{fig:2b}
}
\setcapindent{0em}
\caption{Contour plot for the Higgs matching correction $\delta_H(\mu)$ as a function of the scale $\mu$ and the Higgs mass $m_H$. The left panel shows the Standard Model case which agrees well with previous analyses, see Fig.~1(a) of Ref.~\cite{Hambye:1996wb}. The right panel shows the corrections for the Standard Model with a fourth family of chiral fermions whose masses are indicated in the top line of the graph.}
\label{fig:higgsmatchingconditions}
\end{figure}

\medskip

In \ref{fig:higgsmatchingconditions} we present contour plots for the Higgs matching correction $\delta_H(\mu)$ as a function of the renormalization scale $\mu$ and the Higgs mass $m_H$. The left panel that corresponds to the SM3 case shows excellent agreement with Fig.~(1a) of Ref.~\cite{Hambye:1996wb}. The right panel shows the generalization of $\delta_H(\mu)$ to the case of the SM4.


\section{Generalization of the Top Matching Correction}
\label{sec:topmatchingconditions}

The relation between the top pole mass $m_t$ and its \MS{} mass $\overline{m}_t$ can be expressed as 
\begin{equation}
\overline{m}_t(\mu) = m_t \left( 1+ \delta_t(\mu)\right),
\end{equation}
where the so-called top matching correction is conveniently split into its QCD, QED and weak parts:
\begin{equation}
\delta_t(\mu) = \delta_t^\textrm{QCD}(\mu) + \delta_t^\textrm{QED}(\mu) + \delta_t^\textrm{W}(\mu)
\end{equation}
The QCD and QED corrections are well-known and given by
\begin{equation}
\delta_t^\textrm{QCD}(\mu) = C_F \frac{\alpha_s(\mu)}{4\pi}\left( 3\log\left( m_t^2/\mu^2\right) - 4 \right), \quad \delta_t^\textrm{QED}(\mu) = Q_t^2 \frac{\alpha(\mu)}{4\pi}\left( 3\log\left( m_t^2/\mu^2\right) - 4 \right),
\end{equation}
where $C_F=4/3$ is the quadratic Casimir operator of \SU{3} and $Q_t=2/3$ is the charge of the top quark. The expression for the weak contribution is more involved and reads \cite{Hempfling:1994ar}:
\begin{equation}
\delta_t^\textrm{W}(\mu) = \left.\textrm{Re}\left[ \Sigma^t_V(m_t^2) + \Sigma^t_S(m_t^2) \right] - \frac{\Pi_{WW}^\textrm{bos}(0)+\Pi_{WW}^\textrm{fer}(0)}{2m_W^2} - \frac{E}{2} \enspace \right|_\textrm{finite}
\label{eq:topcorrection}
\end{equation}
$\Sigma^t_V(m_t^2) + \Sigma^t_S(m_t^2)$ is the sum of the vector and scalar components of the top quark self-energy, $E$ is an explicit contribution defined in Ref.~\cite{Sirlin:1985ux}, and $\Pi_{WW}^\textrm{bos}(0)+\Pi_{WW}^\textrm{fer}(0)$ is the sum of the bosonic and fermionic contributions to the $W$ boson self-energy. Explicit expressions for all of these terms can be found in Eq.~(A1), Eq.~(2.8), and Eq.~(A4) of Ref.~\cite{Hempfling:1994ar}, respectively.
\begin{figure}[h!]
\centering
\includegraphics[width=.5\textwidth]{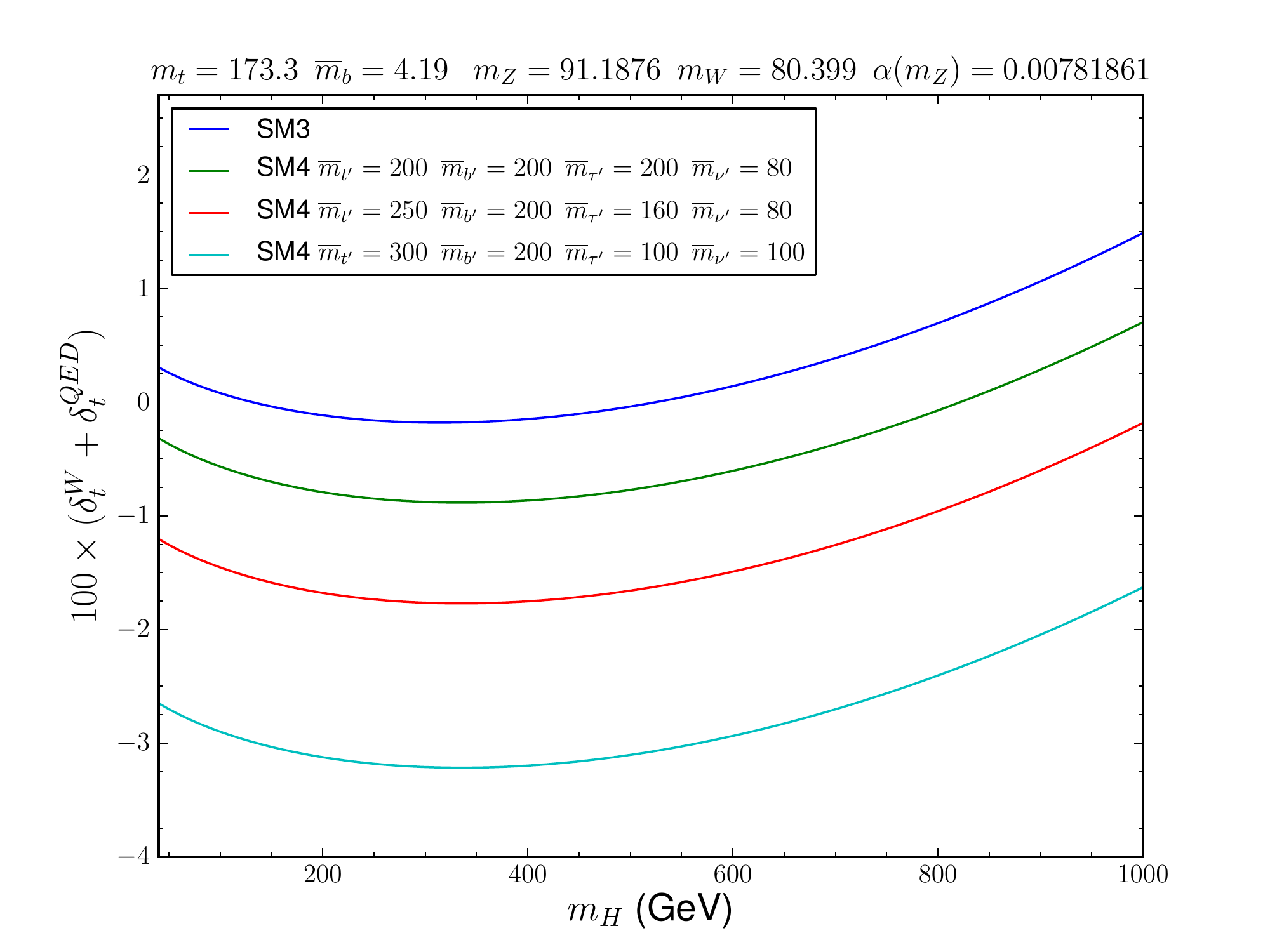}
\setcapindent{0em}
\caption{Electroweak part of the top matching correction $\delta_t(\mu)$ as a function of the Higgs mass $m_H$ (in percentage). The blue curve corresponds to the Standard Model case and agrees reasonably well with the solid line in Fig.~2(a) of Ref.~\cite{Hempfling:1994ar}. The small numerical deviations are likely to stem from the different choice of Standard Model parameters which we have indicated in the top line of the graph.}
\label{fig:topmatchingconditions}
\end{figure}

$\Pi_{WW}^\textrm{fer}(0)$ is the only term in \ref{eq:topcorrection} where the presence of the extra fermions of the fourth generation have an effect and generalizes to:
\begin{align}
\frac{\alpha}{8 \pi  s^2}  &\left[\frac{3 m_b^2 m_t^2 \log \left(m_t^2/m_b^2\right)}{m_t^2-m_b^2}-3 m_b^2
   \left(\frac{1}{2}-\log \left(m_b^2/\mu ^2\right)\right)+\frac{3 m_{b'}^2 m_{t'}^2 \log
   \left(m_{t'}^2/m_{b'}^2\right)}{m_{t'}^2-m_{b'}^2}\right.\notag\\
   & \left.-3 m_{b'}^2 \left(\frac{1}{2}-\log \left(m_{b'}^2/\mu
   ^2\right)\right)+\frac{m_{\tau'}^2 m_{\nu'}^2 \log \left(\frac{m_{\nu'}^2}{m_{\tau'}^2}\right)}{m_{\nu'}^2-m_{\tau'}^2}-m_{\tau'}^2
   \left(\frac{1}{2}-\log \left(\frac{m_{\tau'}^2}{\mu ^2}\right)\right)\right.\\
   & \left.-3 m_t^2 \left(\frac{1}{2}-\log \left(\frac{m_t^2}{\mu
   ^2}\right)\right)-3 m_{t'}^2 \left(\frac{1}{2}-\log \left(\frac{m_{t'}^2}{\mu ^2}\right)\right)-m_{\nu'}^2 \left(\frac{1}{2}-\log
   \left(\frac{m_{\nu'}^2}{\mu ^2}\right)\right)\right].\notag
\end{align}

In \ref{fig:topmatchingconditions} we show the electroweak part of the top matching correction as a function of the Higgs mass $m_H$. We find reasonable agreement with the curve in Ref.~\cite{Hempfling:1994ar} representing the \textit{exact} result. However, we were unable to reproduce the curves corresponding to the \textit{approximate} expressions given by Eq.~(2.15) and Eq.~(2.20) of Ref.~\cite{Hempfling:1994ar}.


\bibliography{mybibliography}

\bibliographystyle{utphys}

\end{document}